\documentstyle[12pt]{article}
\normalbaselines

\def\om{\omega}
\def\ov{\overline}
\def\noi{\noindent}

\begin{document}
\begin{flushright}
UCVFC-DF/9-96
\end{flushright}

\vbox{\vspace{6mm}}

\begin{center}
{\bf A Geometric Representation for the Proca Model}\\ [2mm]
{\it Jaime Camacaro, Rolando Gait\'an and Lorenzo Leal
\footnote{e-mail: lleal@dino.conicit.ve
$\ \ $Postal address: A.P. 47399, Caracas 1041-A, Venezuela}\\ [2mm]
Departamento de F\'{\i}sica, Facultad de Ciencias\\ [2mm]
Universidad Central de Venezuela, Caracas, Venezuela.}
\end{center}

\vspace{1cm}

\begin{abstract}
\noi
The Proca model is quantized in an open-path dependent representation that
generalizes the Loop Representation of gauge theories. The starting point is
a gauge invariant Lagrangian that reduces to the Proca Lagrangian when
certain gauge is selected.
\end{abstract}

\newpage

The quantization of systems with Second-Class Constraints (S.C.C.) has 
received considerable attention in recent times [1-6], mainly due to the 
appearance of this kind of restrictions in super-strings theories [4]. From the point of view of the modern path integral approach, S.C.C. turn the 
quantization procedure into a rather cumbersomed task, which has given rise to 
several interesting developments to handle this situation [7,8].

On the other hand, within the Loop Representation (LR) approach [9-12],
which is the framework we shall adopt in this letter, 
things are not much better. The L.R. was built to quantize gauge-invariant 
theories, which are synonymous of first-class constrained ones. However, one 
feels that there is an underlying geometric content in certain second-class 
constrained theories, that might be accommodated in an adequate L.R. 
formulation.

In this paper we investigate the quantization of the massive Maxwell theory 
(Proca theory) in a geometric representation that generalizes the L.R.. The 
strategy we follow consists on converting the usual Proca theory (which 
posses S.C.C.) into an equivalent gauge-invariant theory that reduces to the 
former when certain gauge is chosen.

This idea is not new. In references [4,5], the S.C.C. of certain theories are 
taken as gauge-fixing conditions of a gauge-invariant theory, whose 
lagrangian is then constructed. Within the same spirit, in references [1-3,6] 
the phase space is enlarged by including aditional variables that turn the 
theory gauge-invariant. Applications of these procedures to concrete models 
are found in references [4,13-18].

In the present case, the model is so simple that there is no need to employ 
any of these methods although the underlying idea we use is the same. We 
start from the usual Proca lagrangian:
$$
{\cal{L}}=-\frac{1}{4}F_{\mu\nu}F^{\mu\nu}+\frac{1}{2}M^2A_\mu A^\mu
\eqno (1)
$$
where $F_{\mu\nu}=\partial_\mu A_\nu -\partial_\nu A_\mu$. The mass term 
spoils the gauge invariance of the first one. To recover this symmetry, we 
add a dinamical auxiliary field $\om$ and consider, instead of (1), the 
lagrangian:
$$
{\cal{L}}'=-\frac{1}{4}F_{\mu\nu}F^{\mu\nu}+\frac{1}{2}M^2(A_\mu +
\partial_\mu \om )(A^\mu +\partial^\mu \om ) \eqno (2)
$$
which is invariant under:
$$
A'_\mu = A_\mu +\wedge_{,\mu} \eqno (3)
$$
$$
\om '=\om - \wedge \eqno (4)
$$
Obviously, there is a gauge choice that gives back (1): $\om$ may be 
eliminated by taking $\wedge =\om$. It is worth mentioning that ${\cal{L}}'$, 
sometimes called the St\"uckelberg lagrangian, has also been obtained from 
(1) through the Batalin, Fradkin and Tyutin method [1,2,18]. ${\cal{L}}'$ can 
also be obtained from the Abelian Maxwell-Higgs model, by ``freezing'' the 
radial mode $|\phi |$ of the Higgs field $\phi =|\phi |e^{i\om}$ and keeping 
only the Goldstone mode $\om$.

The ``pure gauge'' nature of $\om$ is of course also manifested in the 
equations of motion. These are:
$$
M^2(\om^{,\mu}+A^\mu )-F^{\mu\nu}_{,\nu}=0 \eqno (5)
$$
$$
\partial_\mu (\om^{,\mu}+A^\mu )=0 \eqno (6)
$$
Equation (6) is then a consistence condition onto equation (5). In the 
unitary gauge $(\om =0)$ the usual equations of the Proca model are
recovered.

Canonical quatization of ${\cal{L}}'$ may be resumed as follows. There is a 
primary constraint
$$
\pi^0_A \approx 0 \eqno (7)
$$
whose consistency produces the secondary one:
$$
\partial_i \pi^i_A +\pi_\om \approx 0 \eqno (8)
$$
Here, $\pi^\mu_A$ and $\pi_\om$ are the canonical conjugates to $A^\mu$ and 
$\om$ respectively. The Hamiltonian results to be:
$$
H=\int d^3x \left\{\frac{1}{2}(\pi^i_A)^2+\frac{1}{2M^2}\pi^2_\om +\frac{1}{4}
(F_{ij})^2+\frac{M^2}{2}(\om_{,i}+A_i)^2+A_0(\pi_\om +\pi^i_{A,i})\right\}
\eqno (9)
$$
and the non vanishing canonical commutators are given by:
$$
[A_\mu (y),\pi^\nu_A(x)]=i\delta^\nu_\mu \delta^3(x-y) \eqno (10.a)
$$
$$
[\om (x),\pi_\om (y)]=i\delta^3(x-y) \eqno (10.b)
$$
It is easily verified that do not appear further constraints, and that (7) 
and (8) are first class ones, as corresponds to a gauge-invariant theory. To 
recover the usual canonical formulation of the Proca model, one should fix 
the gauge by adding the conditions:
$$
\om \approx 0 \eqno (11.a)
$$
$$
\pi_\om +M^2A_0 \approx 0 \eqno (11.b)
$$
which, together with (7), (8) would complete a system of S.C.C.. Then, a 
straightforward calculation, which includes the obtention of the appropriate 
Dirac brackets yields the standard result.

Returning to the gauge invariant formulation, we see that the primary 
constraint (7) states that on the physical subspace, the wave functional
$\Psi (\om ,A_\mu )$ do not dependent on $A_0$ (we have momentarily chosen
the Schr\"odinger polarization). Thus, operators $A_0$ and $\pi_0$ may be
ignored. Alternatively, one may choose the temporal gauge $A_0\approx 0$,
which together with eq. (7) leads to the Dirac commutator $[A_0,\pi_0]=0$.
Then, both $A_0$ and $\pi_0$ may be taken as strongly vanishing operators.
Regarding the ``Gauss constraint'' (8), it generates the time-independent
gauge transformations of the theory. It is inmediate to see that both
$\om$ and $A_i$ are the gauge-dependent operators, unlike their canonical
conjugates $\pi_\om$ and $\pi^i$.

Within the spirit of the L.R. formulation, we introduce a
gauge-invariant non local operator that encodes the physical information
associated to $\om$ and $A_i$. A natural candidate is:
$$
f(P_x^y)=exp(-i\om (x))W(P_x^y)exp(i\om (y))\eqno (12)
$$
where
$$
W(P_x^y)=exp(i\int_{P_x^y}dz^kA_k(z))\eqno (13)
$$
is the Wilson-path operator associated to the spatial path $P_x^y$ starting
at x and ending at y. More generally, define a ``generalized path''
${\bf{\cal{P}}}$ as an unordered colection of open $(P_x^y)$ and closed
$(P_x^x)$ paths
$$
{\bf{\cal{P}}}\equiv \left\{P_{x_1}^{(1)}{}^{y_1},\cdots,
P_{x_N}^{(N)}{}^{y_N}\right \} \eqno (14)
$$
From equation (12), we define:
$$
f({\bf{\cal{P}}})\equiv \prod_{\alpha =1}^N f(P^{(\alpha )})\eqno (15)
$$
It is easy to see that
$$
f(\ov{{\bf{\cal{P}}}})f({\bf{\cal{P}}})={\bf{1}} \eqno (16)
$$
where ${\bf{\ov{\cal{P}}}}=\left\{\ov{P}^{(1)},\cdots ,\ov{P}^{(N)}\right\}$
is the set of paths opposite to $\ov{P}^{(1)},\cdots ,\ov{P}^{(N)}$. Equation
(16) allows to build a group of generalized paths (GGP) that mimics the
Abelian Group of Loops of Gambini-Tr\'{\i}as [9]. We define the elements of
the GGP by grouping the generalized paths ${\bf{\cal{P}}}$ into equivalence
classes given by:
$$
{\bf{\cal{P}}}\sim {\bf{\cal{Q}}}\ \ \ \ iff \ \ \ \
f(\ov{{\bf{\cal{P}}}})f({\bf{\cal{Q}}})={\bf{1}} \eqno (17)
$$
Then, it can be seen that the operation:
$$
{\bf{\cal{P}}} \cdot {\bf{\cal{Q}}} \equiv \left \{P^{(1)},\cdots ,P^{(N)},
Q^{(1)},\cdots ,Q^{(M)}\right \}\eqno (18)
$$
defines a group product among the equivalence classes. It is worth mentioning
that this group structure is absent in Scalar Electrodynamics, although even
in this case the analogous to $f(P_x^y)$ (eq. (11)) is a gauge-independent
operator.

The algebra obeyed by $\pi_\om$, $\pi_A^i$ and $f({\bf{\cal{P}}})$ is
given by:
$$
\left [ \pi_\om (z),f({\bf{\cal{P}}})\right ] =\sum_{\alpha =1}^N
\rho (z,P_{x_\alpha}^{y_\alpha})f({\bf{\cal{P}}})\eqno (19.a)
$$
$$
\left [ \pi_A^i(z),f({\bf{\cal{P}}})\right ] =\sum_{\alpha =1}^N
T^i (z,P_{x_\alpha}^{y_\alpha})f({\bf{\cal{P}}})\eqno (19.B)
$$
where
$$
\rho (z,{\bf{\cal{P}}}_{x_\alpha}^{y_\alpha})=\delta^3(z-y_\alpha )
-\delta^3(z-x_\alpha )\eqno (20)
$$
and $T^i(z,P)$ is the distributional field of tangent vectors associated to
the path $P$:
$$
T^i(z,P_x^y)=\int_{P_x^y}dx^{'i}\delta^3(x'-z)\eqno (21)
$$
This algebra may be realized onto path-dependent wave functionals
$\Psi({\bf{\cal{P}}})$ by prescribing:
$$
\pi_\om (z) \Psi({\bf{\cal{P}}}) =\sum_{\alpha =1}^N
\rho(z,P_{x_\alpha}^{y_\alpha})\Psi({\bf{\cal{P}}}) \eqno (22)
$$
$$
\pi_A^i (z) \Psi({\bf{\cal{P}}}) =\sum_{\alpha =1}^N
T^i(z,P_{x_\alpha}^{y_\alpha})\Psi({\bf{\cal{P}}}) \eqno (23)
$$
$$
f({\bf{\cal{Q}}})\Psi({\bf{\cal{Q}}})=\Psi({\bf{\cal{Q}}}\cdot{\bf{\cal{P}}})
\eqno (24)
$$
Since
$$
\frac{\partial}{\partial z^i}T^i(z,P_{x\alpha}^{y_\alpha}) +
\rho(z,P_{x\alpha}^{y_\alpha})=0 \eqno (25)
$$
the Gauss constraint (eq. (8)) is automatically verified.

Equations (22)-(24) admit an easy geometrical interpretation. The operator
$f({\bf{\cal{P}}})$performs ``translations'' on the generalized path space,
while $\pi_\om$ and $\pi_A^i$ measure the density of pairs of points and
the ``shape'' of the lines, respectively, associated to the paths. From a
physical point of view, it can be said that in the
${\bf{\cal{P}}}$-representation, $f({\bf{\cal{P}}})$ creates lines of
electric field with Goldstone bosons of opposite signs attached at its end
points, $\pi_\om$ measures the density of such bosons and $\pi_A^i$ measures the density of electrical flux.

Next we show how to write the Hamiltonian $H$ in the
${\bf{\cal{P}}}$-representation. To this end, we introduce a slightly
modified Mandelstam derivative [19] $\Delta_k(z)$, defined as follows. Given
a functional $G({\bf{\cal{P}}})$ depending on generalized paths
${\bf{\cal{P}}}$, $\Delta_k(z)G({\bf{\cal{P}}})$ computes the change of $G$
when an infinitesimal open path $\delta P_z^{z+h}(h\to 0)$ starting at $z$
and ending at $z+h$ is appended to the list of paths $P^{(\alpha)}$ comprised
in ${\bf{\cal{P}}}$:
$$
\Delta_iG({\bf{\cal{P}}})=\lim_{h^i\to 0} \frac{G(\delta P,P^{(1)},\cdots ,
P^{(N)})-G(P^{(1)},\cdots ,P^{(n)})}{h^i}\eqno (26)
$$
The $\Delta_k(z)$ derivative is related with the ``area'' or ``loop''
derivative $\Delta_{ij}(z)$ of Gambini-Tr\'{\i}as [9] through the expresion:
$$
\Delta_{ij}(z)=\frac{\partial}{\partial z^i}\Delta_j(z) -
\frac{\partial}{\partial z^j}\Delta_i(z) \eqno (27)
$$
The area derivative measures how a path-dependent functional changes when a
small ``plaquette'' is attached at the point $z$.

In terms of these geometric derivatives, the stationary Schr\"o\-din\-ger
equation in the ${\bf{\cal{P}}}$-re\-pre\-sen\-ta\-tion looks as:

$$
\int d^3z \left\{ -\Delta_{ij}(z)\Delta_{ij}(z) - \frac{M^2}{2}\Delta_k(z)
\Delta_k(z) + \frac{1}{2}T^i(z,{\bf{\cal{P}}}) T^i(z,{\bf{\cal{P}}}) + 
\frac{M^2}{2}\rho^2(z,{\bf{\cal{P}}})\right\}\Psi({\bf{\cal{P}}})
=E\Psi({\bf{\cal{P}}})\eqno (28)
$$

where we have set:
$$
T^i(z,{\bf{\cal{P}}})=\sum^N_{\alpha =1}T^i(z,P_{x_\alpha}^{y_\alpha})
\eqno (29.a)
$$
$$
\rho (z,{\bf{\cal{P}}})=\sum^N_{\alpha =1}\rho (z,P_{x_\alpha}^{y_\alpha})
\eqno (29.b)
$$
The left hand side of equation (28) has the following structure. It comprises
a ``kinetic energy'' term, constructed from generalized laplacians
$(\Delta_{ij})^2$ and $(\Delta_i)^2$ acting on path-space. The remaining
terms may be viewed as ``potential energy'' contributions, that are quadratic
in the ``position variables'' $T^i(z,{\bf{\cal{P}}})$,
$\rho (z,{\bf{\cal{P}}})$, associated to the generalized path
${\bf{\cal{P}}}$. This structure generalizes that of the Maxwell case in the
LR [10], with which our formulation should be compared.

As in the massless case [10], the Schr\"odinger equation (28) only holds at
formal level, in the sense that involves ill defined products of
distributions that should be appropriately regularized. Nevertheless, even at
a formal level, it is possible to solve the functional eigenvalue problem
(26) to obtain the ${\bf{\cal{P}}}$-dependent eigenfunctions and their
corresponding energies, starting from a formal vacuum functional of the type
$\Psi_0({\bf{\cal{P}}})=exp(-\frac{1}{2}\int_{{\bf{\cal{P}}}}dx^i
\int_{{\bf{\cal{P}}}}dy^i D_{ij}(x-y))$ and using standard Fourier methods.

It seems possible to extend the above formulation to treat the more
interesting non Abelian massive case. The so called $2+1$ ``self-dual''
theory [20], which consists of a Chern-Simons term plus a mass term, could
also be formulated in an appropriate path representation, following the
present scheme of quantization.


\newpage

\begin{thebibliography}{99}
\bibitem{}I.A. Batalin and E.S. Fradkin, Nucl. Phys. {\bf B279} (1987) 514;
Phys. Lett. {\bf B180} (1986) 157.
\bibitem{}I.A. Batalin and J.V. Tyutin, Int. J. Mod. Phys. {\bf A6} (1991)
3255.
\bibitem{}A. Restuccia and J. Stephany, Phys. Lett {\bf B305} (1993) 348.
\bibitem{}R. Gianvittorio, A. Restuccia and J. Stephany, Mod. Phys. Lett.
{\bf A6} (1991) 2121.
\bibitem{}K. Harodo and H. Mukaida, Z. Phys. {\bf C48} (1990) 151.
\bibitem{}I.A. Batalin, E.S. Fradkin and T.E. Fradkina, Nucl. Phys. {\bf 314}
(1989) 158.
\bibitem{}M.B. Green and J. Schwarz, Phys. Lett. {\bf B109} (1983) 399.
\bibitem{}L. Brink and J. Schwarz, Phys. Lett. {\bf B100} (1981) 287.
\bibitem{}R. Gambini and A. Tr\'{\i}as, Phys. Rev. {\bf D22} (1980) 1380;
ibid {\bf D23} (1981) 553; ibid {\bf D27} (1983) 2935; ibid {\bf D31} (1985)
3144; Nucl. Phys. {\bf B278} (1986) 436.
\bibitem{}C. Di Bartolo, F. Nori, R. Gambini and A. Tr\'{\i}as, Lett. Nuovo
Cim. {\bf 38} (1983) 497.
\bibitem{}R. Gambini. L. Leal and A. Tr\'{\i}as, Phys. Rev. {\bf D39} (1989)
3127.
\bibitem{}C. Rovelli and L. Smolin, nucl. Phys. {\bf 331} (1990) 80.
\bibitem{}M. Moshe and Y. Oz, Phys. Lett. {\bf B224} (1989) 145.
\bibitem{}T. Fujiwara, Y. Igarashi and J. Kubo, Nucl. Phys. {\bf B341} (1990)
695.
\bibitem{}J. Kowalski-Glikman, Phys. Lett. {\bf B245} (1990) 79.
\bibitem{}I.A. Batalin and I.V. Tyutin, Mod. Phys. Lett. {\bf A7} (1992)
1931.
\bibitem{}N. Banerjee, Subir Ghosh and R. Banerjee, Phys. Rev. {\bf D49}
(1996) 49.
\bibitem{}N. Banerjee and R. Banerjee, {\it Generalized Hamiltonian embedding
of the Proca model}, S.N. Bose National Centre for Basic Sciencies-preprint
(1996).
\bibitem{}S. Mandelstam, Ann. Phys. {\bf 19} (1962) 1.
\bibitem{}P.K. Townsend, K. Pilch and P. van Nieuwenhuizen, Phys. Lett.
{\bf 136} (1984) 38.
\end{thebibliography}
\end{document}